\documentclass[pdflatex,sn-mathphys-num]{sn-jnl}
\usepackage{color,soul}
\usepackage{multirow}%
\usepackage{amsmath,amssymb,amsfonts}%
\usepackage{amsthm}%
\usepackage{mathrsfs}%
\usepackage[title]{appendix}%
\usepackage{xcolor}%
\usepackage{textcomp}%
\usepackage{manyfoot}%
\usepackage{booktabs}%
\usepackage{algorithm}%
\usepackage{algorithmicx}%
\usepackage{algpseudocode}%
\usepackage{listings}%
\usepackage{todonotes}
\usepackage{graphicx}%
\usepackage{comment}
\usepackage{subfigure}

\begin{document}

\title{Plug-and-Play with 2.5D Artifact Reduction Prior for Fast and Accurate Industrial Computed Tomography Reconstruction}

\author*[1]{\fnm{Haley} \sur{Duba-Sullivan}}\email{sullivanhe@ornl.gov}

\author[2]{\fnm{Aniket} \sur{Pramanik}}\email{aniketpramanik16@gmail.com}

\author[3]{\fnm{Venkatakrishnan} \sur{Singanallur}}\email{venkatakrisv@ornl.gov}

\author*[3]{\fnm{Amirkoushyar} \sur{ Ziabari}}\email{ziabarikak@ornl.gov}

\affil[1]{\orgdiv{Cyber Resilience and Intelligence Division}, \orgname{Oak Ridge National Laboratory}, \orgaddress{\street{One Bethel Valley Road}, \city{Oak Ridge}, \postcode{37831}, \state{TN}, \country{USA}}}

\affil[2]{\orgname{Memorial Sloan Kettering Cancer Center}, \orgaddress{\street{411 E 75th St} \city{New York}, \postcode{10021}, \state{NY}, \country{USA}}}

\affil[3]{\orgdiv{Electrification and Energy Infrastructures Division}, \orgname{Oak Ridge National Laboratory}, \orgaddress{\street{One Bethel Valley Road}, \city{Oak Ridge}, \postcode{37831}, \state{TN}, \country{USA}}}

\abstract{Cone-beam X-ray computed tomography (XCT) is an essential imaging technique for generating 3D reconstructions of internal structures, with applications ranging from medical to industrial imaging. 
Producing high-quality reconstructions typically requires many X-ray measurements; this process can be slow and expensive, especially for dense materials. 
Recent work incorporating artifact reduction priors within a plug-and-play (PnP) reconstruction framework has shown promising results in improving image quality from sparse-view XCT scans while enhancing the generalizability of deep learning-based solutions. 
However, this method uses a 2D convolutional neural network (CNN) for artifact reduction, which captures only slice-independent information from the 3D reconstruction,
limiting performance. 
In this paper, we propose a PnP reconstruction method that uses a 2.5D artifact reduction CNN as the prior. 
This approach leverages inter-slice information from adjacent slices, capturing richer spatial context while remaining computationally efficient. 
We show that this 2.5D prior not only improves the quality of reconstructions but also enables the model to directly suppress commonly occurring XCT artifacts (such as beam hardening), eliminating the need for artifact correction pre-processing. Experiments on both experimental and synthetic cone-beam XCT data demonstrate that the proposed method better preserves fine structural details, such as pore size and shape, leading to more accurate defect detection compared to 2D priors. 
In particular, we demonstrate strong performance on experimental XCT data using a 2.5D artifact reduction prior trained entirely on simulated scans, highlighting the proposed method’s ability to generalize across domains.
}


\maketitle

\section{Introduction}\label{sec:introduction}

X-ray computed tomography (XCT) is a critical technique with many applications in medical and industrial imaging. 
In industrial XCT, reconstructing a 3D object requires solving a large inverse problem using many projections from different angles.
The resulting reconstruction can be used for internal inspection and anomaly detection, with applications in additive manufacturing (e.g. nondestructive evaluation) and medical imaging (e.g. tumor detection)
\cite{azhari2014tumor,zanini2017assembly,rupal2020geometric,khosravani2020use,naresh2020use,  liu2024universal}. 
However, attaining high-quality reconstructions can be difficult; the quality of the reconstruction depends on factors such as desired reconstruction resolution, total integration time, total number of views, and X-ray scan setting (e.g. voltage, current, and physical filters). 
Higher quality generally requires more time and higher radiation doses. 

Traditional methods such as Feldkamp, Davis and Kress (FDK)~\cite{fdk} can quickly produce reconstructions, but require a large number of projections at sufficiently high signal-to-noise ratio to achieve high quality, resulting in longer scans with more exposure.
More advanced algorithms such as model-based iterative reconstruction (MBIR) methods~\cite{bouman1993generalized,sabo92tv,thibault2007three,yu2010fast,venkatakrishnan2021algorithm} and plug-and-play (PnP) methods~\cite{venkatakrishnan2013plug,chan2016plug, he2018optimizing, kamilov2023plug, vo2024plug}, can produce high-quality reconstructions even with sparse measurements, but they are slow and computationally expensive.
Deep learning has been proposed as a fast alternative that directly maps low-quality reconstructions to high-quality reconstructions~\cite{hammernik2017deep, park2018ct, xu2018deep, ghani2019fast, ziabari2023enabling}.
However, these models rely on large amounts of representative training data and often struggle to generalize to new XCT scans that differ in part geometry, material composition, print parameters, or scan settings. 
As a result, their performance can degrade significantly when applied to data that fall outside the distribution of the training set.

To address these challenges, a practical PnP algorithm that is both flexible and efficient was proposed in~\cite{pramanik2025fast}. 
This method uses an artifact reduction CNN prior instead of a CNN-based Gaussian denoiser, along with an adaptive regularization parameter selection strategy to perform reconstruction for large scale 3D imaging data. 
However, the artifact reduction CNN prior only exploits single-slice information from the 3D reconstruction, which may limit its performance.

2.5D deep learning architectures leverage multi-slice information by aggregating neighbouring slices along the channel dimension to estimate the center slice.
The ability of 2.5D architectures to produce better image quality than 2D architectures has been established in a variety of applications, such as XCT super-resolution~\cite{duba20242}, XCT image denoising~\cite{2.5Ddenoising}, volumetric image segmentation~\cite{2.5Dsegmentation}, and XCT image reconstruction~\cite{majee2021multi, luo20162,2.5DMBIR,rahman2023mbir,ziabari2023enabling, ziabari2022simurgh}.
2.5D architectures have the added benefit of maintaining low computational complexity in contrast to 3D architectures. {Prior work in volumetric imaging~\cite{angermann2019projection, 2.5DMBIR, zhang2022bridging, he2018optimizing} similarly supports the use of 2.5D models as efficient alternatives to 3D networks, especially in high-resolution domains like XCT.}
In this paper, we propose an improved practical PnP method that uses a 2.5D artifact reduction prior which incorporates multi-slice information from the 3D reconstruction while preserving low computational complexity.
This simple modification improves reconstruction quality without the heavy cost of 3D models.
We present results on experimental and synthetic cone-beam XCT datasets that demonstrate improved performance on both in-distribution (InD) and out-of-distribution (OOD) data when using a 2.5D artifact reduction prior instead of a 2D prior.
In particular, we demonstrate strong performance on experimental XCT data using a model trained entirely on synthetic scans, highlighting the method's ability to generalize across domains.

\begin{figure}
    \centering
    \includegraphics[width=\linewidth]{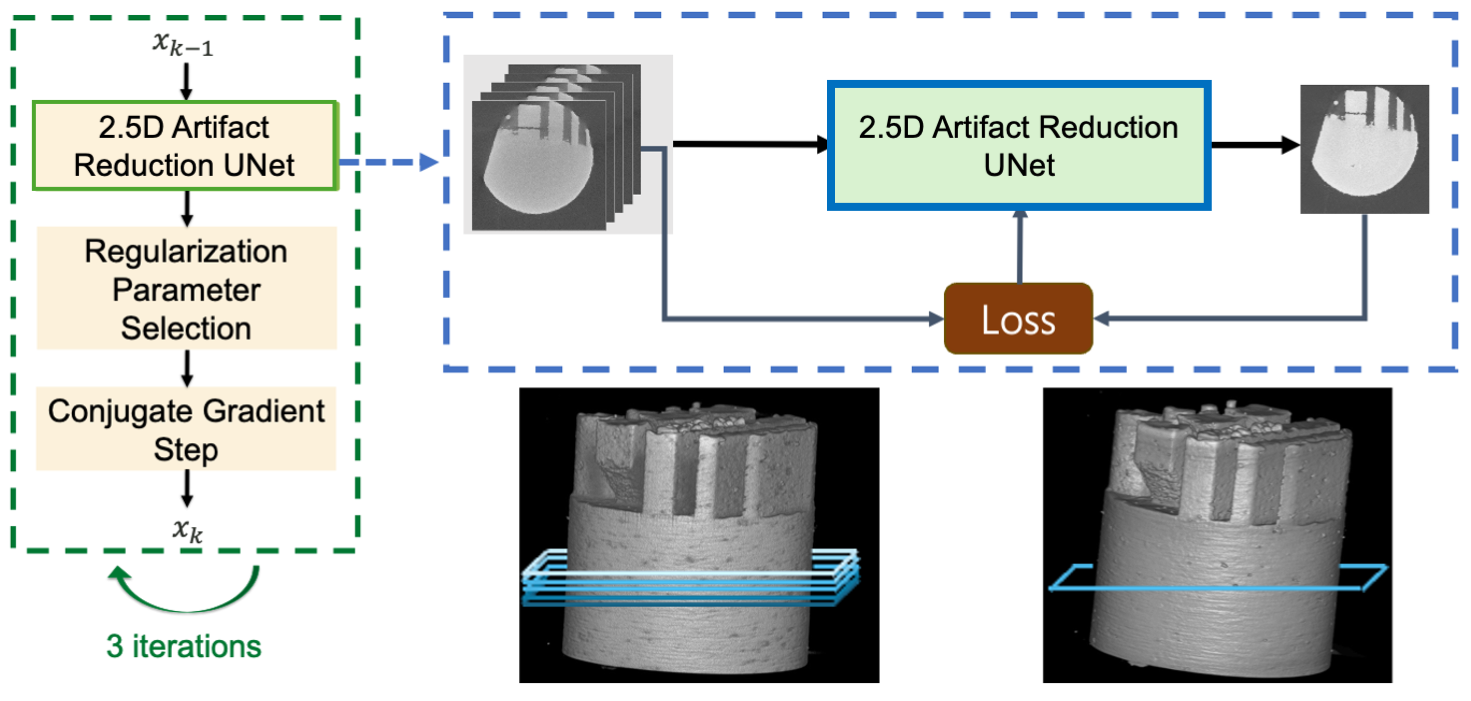}
    \caption{Pipeline of our proposed 2.5D PnP method. 
    The 2.5D artifact reduction prior takes 5 neighboring slices from a sparse-view FDK reconstruction containing beam hardening artifacts as input and reduces both noise and artifacts from the center slice in order to match the dense-view FDK reconstruction that does not contain beam hardening artifacts. 
    }
    \label{fig:pipeline}
\end{figure}

\begin{table*}
\fontsize{8}{11}
\selectfont
\centering
\begin{tabular}{|c|ccccc|}
\hline
\multicolumn{6}{|c|}{\textbf{Synthetic Cone-beam XCT Aluminum Datasets}} \\ \hline
\multirow{2}{*}{\textbf{Dataset}} & \multirow{2}{*}{\textbf{Sample No.}} & \textbf{\# of views} & \textbf{Short-Scan} & \textbf{Noise Level} & \textbf{Beam Hardening}\\ 
 & & [Input, Ref.] & [Input, Ref.] & [Input, Ref.] & [Input, Ref.] \\ \hline 
\multirow{6}{*}{Training Set} & 1 &[580, 2132] & [Yes, No] & [0.0, 0.0] & [Yes, No]\\
 & 1 & [580, 2132] & [Yes, No] & [0.5, 0.0]& [Yes, No]\\
 & 1 & [580, 2132] & [Yes, No] & [1.0, 0.0]& [Yes, No]\\
 & 1 & [145, 2132] & [Yes, No] & [0.0, 0.0]& [Yes, No]\\
 & 1 & [145, 2132] & [Yes, No] & [0.5, 0.0]& [Yes, No]\\
 & 1 & [145, 2132] & [Yes, No] & [1.0, 0.0]& [Yes, No]\\
 \hline
 \multirow{6}{*}{InD Test Set} & 2 & [580, 2132] & [Yes, No] & [0.0, 0.0] & [Yes, No]\\
 & 2 & [580, 2132] & [Yes, No] & [0.5, 0.0]& [Yes, No]\\
 & 2 & [580, 2132] & [Yes, No] & [1.0, 0.0]& [Yes, No]\\
 & 2 & [145, 2132] & [Yes, No] & [0.0, 0.0]& [Yes, No]\\
 & 2 & [145, 2132] & [Yes, No] & [0.5, 0.0]& [Yes, No]\\

 & 2 & [145, 2132] & [Yes, No] & [1.0, 0.0]& [Yes, No]\\
\hline
 \multirow{7}{*}{OOD Test Set} 
 & 2 & [580, 2132] & [Yes, No] & [0.75, 0.0]& [Yes, No]\\ 
 & 3 & [580, 2132] & [Yes, No] & [2.0, 0.0] & [Yes, No]\\
 & 3 & [580, 2132] & [Yes, No] & [4.0, 0.0] & [Yes, No]\\
 & 2 & [145, 2132] & [Yes, No] & [0.75, 0.0]& [Yes, No]\\
 & 3 & [145, 2132] & [Yes, No] & [2.0, 0.0] & [Yes, No]\\
 & 3 & [73, 2132] & [Yes, No] & [0.5, 0.0] & [Yes, No]\\
 & 3 & [73, 2132] & [Yes, No] & [1.0, 0.0] & [Yes, No]\\
\hline
\end{tabular}
\caption{Synthetic cone-beam XCT aluminum datasets used to evaluate 2D and 2.5D PnP. 
The In-distribution data (InD) test set contains scans which match the training data in the material scanned, number of views, presence of beam hardening, and noise levels. 
The out-of-distribution (OOD) test set contains scans where the noise level or number of views is different than those used to train the CNNs.
}
\label{tab:ct_data} 
\end{table*}

\section{Plug-and-Play with 2.5D Artifact Reduction Prior}

The forward model for a cone-beam XCT system is given by $y=Ax$ where $y \in \mathbb{R}^{M}$ contains the projection measurements, $A \in \mathbb{R}^{M\times N}$ is a linear operator encoding the cone-beam projection, and $x\in \mathbb{R}^N$ is the 3D volume of linear attenuation coefficients that we would like to reconstruct. 
A common approach for estimating $x$ is to use a regularized weighted least-squares formulation, i.e.,
\begin{equation}\label{eq:mbir}
    \hat{x} = \arg \min_{x} \left \{ \frac{1}{2} \|Ax - y\|_2^2 + \lambda R(x) \right \},
\end{equation}
where $R(x)$ is a regularizer that encourages certain ``desirable'' properties in the reconstruction and $\lambda$ is a parameter weighting the impact of the regularizer.
{We solve this minimization problem using a quadratic penalty method with alternating minimization, as is done in \cite{pramanik2025fast}. Namely, we introduce an auxiliary variable $z$ to decouple the regularizer:
\begin{equation}
    \hat{x}, \hat{z} = \arg \min_{x, z} \left \{ \frac{1}{2} \|Ax - y\|_2^2 + \lambda R(z) \right \} \text{ such that } x=z.
\end{equation}
Instead of directly enforcing this constraint, we relax it using a quadratic penalty term with tunable parameter $\beta > 0$:
\begin{equation}\label{eq:relaxed_mbir}
    \hat{x}, \hat{z} = \arg \min_{x, z} \left \{ \frac{1}{2} \|Ax - y\|_2^2 + \lambda R(z) + \frac{\beta}{2} \|x-z\|^2 \right \}.
\end{equation}
We then solve the relaxed minimization problem in~\eqref{eq:relaxed_mbir} using alternating minimization, which} provides an iterative algorithm that alternates between a data-fitting sub-problem and a regularization sub-problem, i.e.,
\begin{align}
    \hat{z}_k &= \arg \min_{z} \left \{ \lambda R(z) + \frac{\beta}{2} \|\hat{x}_{k-1} - z\|_2^2 \right \}~\label{eq:reg_prox}\\
    \hat{x}_k &= \arg \min_{x} \left \{  \frac{1}{2} \|Ax - y\|_2^2 + \frac{\beta}{2} \|x - \hat{z}_k\|_2^2 \right \}.~\label{eq:data-fitting_prox}
\end{align}
{The proposed algorithm can be interpreted as a proximal splitting scheme with a quadratic penalty, and it is closely related to majorization-minimization and variable-splitting methods. Note that for the constraint $x=z$ to be enforced, we must increase $\beta$ as $k\to \infty$. Rather than setting an explicit schedule for $\beta$, we use  the proposed adaptive parameter selection strategy from~\cite{pramanik2025fast} to update $\beta$ at each iteration, which selects $\beta$ using a grid search based on the reconstruction quality of a few center slices from the 3D volume. Pseudocode and more detail for this parameter selection algorithm can be found in \cite{pramanik2025fast}.}

Due to the large size of $A$, it is not trivial to solve \eqref{eq:data-fitting_prox}. 
Instead, it is common to use a few iterations of either the conjugate gradient method (CGM) or a gradient descent-based method to estimate $\hat{x}_k$. 
To that end, we use CGM for this step.

{PnP methods build upon the insight that the regularization sub-problem in \eqref{eq:reg_prox} can be interpreted as a maximum a posteriori (MAP) estimate in a Gaussian denoising problem~\cite{buzzard2018consensus}. This equivalence allows the proximal operator associated with the prior to be replaced by an off-the-shelf denoiser, typically a CNN trained on problem-specific data. The PnP framework thus decouples the data fidelity and prior modeling steps, enabling flexible incorporation of powerful learned denoisers without explicitly formulating a regularizer. Convergence properties of PnP algorithms have been studied extensively, with results often relying on denoisers satisfying nonexpansiveness or averaged operator conditions~\cite{nair2021fixed, buzzard2018consensus}. In practice, these methods have demonstrated impressive empirical performance across various inverse problems, including imaging reconstruction, where handcrafted priors are insufficient or difficult to specify.}

Instead {of using a generic denoiser}, here we use a CNN trained to reduce artifacts from sparse-view and noisy FDK reconstructions. 
As for the choice of the artifact reduction prior, we propose to use a 2.5D architecture that allows for exploiting multi-slice information to more effectively reduce artifacts and noise while preserving the underlying structure of the data. 
{Compared to full 3D networks, our 2.5D design offers a favorable trade-off between computational cost and performance; it captures important inter-slice context while remaining significantly more memory-efficient and easier to train.
This makes 2.5D particularly well-suited for PnP methods applied to large volumetric datasets.}
Figure~\ref{fig:pipeline} shows a high-level overview of our proposed method, and Section~\ref{sec:2.5d_arch} provides a more detailed discussion of our 2.5D artifact reduction prior.
{Additionally we provide pseudocode for our proposed 2.5D PnP algorithm in Algorithm~\ref{alg:pnp_alg}.}

\begin{algorithm}[t!]
  \caption{{Proposed 2.5D Artifact Reduction PnP Algorithm}}
  \label{alg:pnp_alg}
  \begin{algorithmic}
  \State \textbf{Input:} low-quality FDK reconstruction $\mathbf x_0 = \mathbf x_{\rm FDK}$, $\mathbf A$, $\mathbf A_c$, $\mathbf y$, $K$
  \State \textbf{Output:} high-quality reconstruction $\widehat{\mathbf x}$
  \State $\mathbf y_c \leftarrow$ Measurements from center rows of $\mathbf y$
  \For{$k = 1, \dots K$}
    \State $\mathbf z_k \leftarrow \mathbf D_{\theta}(\mathbf x_{k-1}) \hspace{5.0cm}\textit{\# 2.5D Artifact Reduction Prior}$
    \State $\beta_k \leftarrow$ \textit{RegularizationSelection}($\mathbf z_k$, $\mathbf A_c$, $\mathbf y_c$)
    \State $\mathbf x_k \leftarrow \arg \displaystyle\min_{\mathbf x} \left\{ \frac{1}{2}\|\mathbf A \mathbf x - \mathbf y\|_2^2 + \frac{\beta_k}{2} \|\mathbf x - \mathbf z_k\|_2^2 \right\}$ \hspace{0.25cm}\textit{\# Conjugate Gradient Method}
  \EndFor \\
  \Return $\widehat{\mathbf x} = \mathbf x_K$
  \end{algorithmic}
\end{algorithm}

\section{Implementation Details}

In this section, we outline the datasets used for our experimental results as well as the implementation of our 2.5D artifact reduction CNN {and the hyperparameters used.}

\subsection{2.5D Artifact Reduction CNN}\label{sec:2.5d_arch}
\begin{figure}
    \centering
    \includegraphics[width=\linewidth]{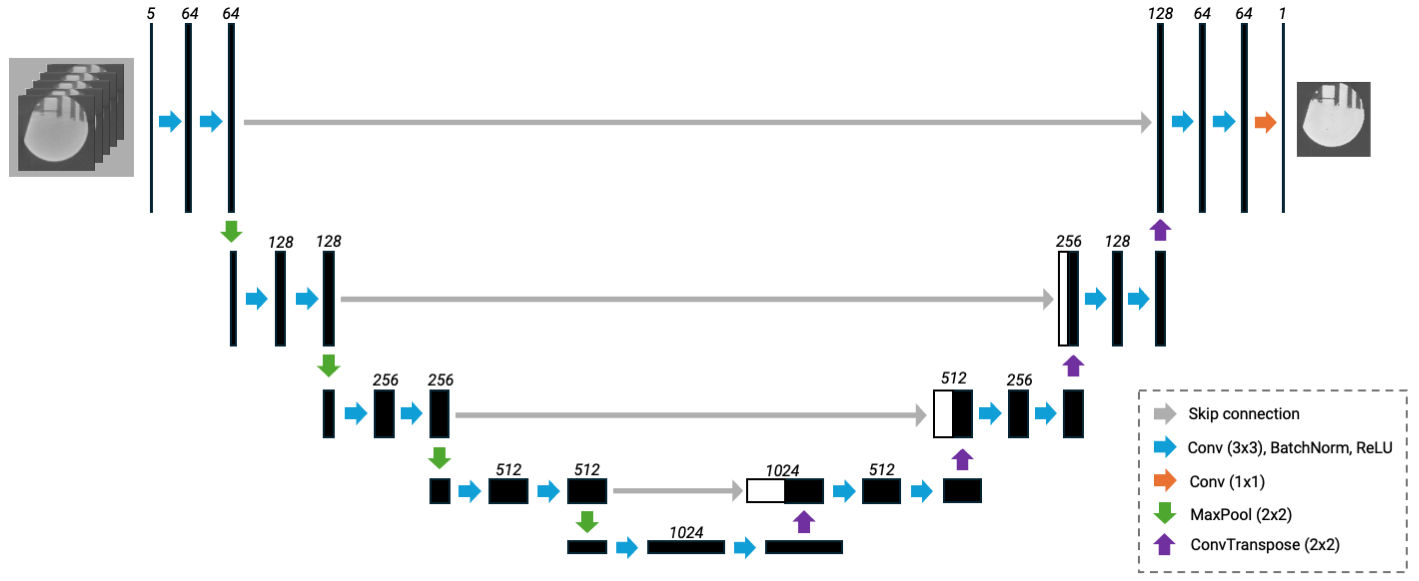}
    \caption{{Architecture of our 2.5D artifact reduction UNet used as the prior model in the proposed 2.5D PnP algorithm.}}
    \label{fig:unet_arch}
\end{figure}

In contrast with other methods that use artifact reduction priors~\cite{liu2020rare, hu2023restoration, pramanik2025fast}, we propose to use a 2.5D architecture for our artifact reduction prior.
Namely, we use the UNet architecture from~\cite{ronneberger2015u}, which consists of four pooling/unpooling layers. 
However, instead of providing one slice of the reconstruction as input to the network, we provide a stack of 5 neighboring slices from a sparse-view FDK reconstruction containing BH artifacts as input, modifying the number of channels in the first layer of the UNet accordingly. 
{Figure~\ref{fig:unet_arch} shows the architecture of our 2.5D UNet.} 
The 2.5D artifact reduction network learns to reduce noise and artifacts from the center slice in order to match a dense-view FDK reconstruction that does not contain BH artifacts.
This modification allows the CNN to learn multi-slice information from the 3D volume, rather than only single-slice information. 

We divide the training volumes into patches of size $5\times 256\times 256$, and split these into training and validation patches with a 80/20 ratio respectively.
Then, we train the 2.5D UNet for 200 epochs with an Adam optimizer~\cite{kingma2014adam}. 
We initialize the learning rate at $1 \times 10^{-3}$, and reduce it by a factor of 2 when the normalized root mean square error (NRMSE) of the validation patches stops improving for 10 epochs. 
After training, we use the epoch that attains the lowest NRMSE on the validation data for all further testing.
For comparison, we also train a 2D UNet with the same training procedure. 

The network is trained by minimizing an L1 loss defined as:
\begin{equation}
    \mathcal{L}(\boldsymbol{\theta}) = \frac{1}{N} \sum_{i=1}^{N} \left| R(\boldsymbol{x}_i; \boldsymbol{\theta}) - \left( \mathcal{P}_{\text{c}} \boldsymbol{x}_i - \boldsymbol{y}_i \right) \right|,
\end{equation}
where $\boldsymbol{x}_i \in \mathbb{R}^{M \times M \times 5}$ is an input patch consisting of five adjacent slices, and $\boldsymbol{y}_i \in \mathbb{R}^{M \times M}$ denotes the corresponding target center slice. The operator $\mathcal{P}_{\text{c}}$ extracts the center slice from the input stack, and $R(\cdot; \boldsymbol{\theta})$ is the network with parameters $\boldsymbol{\theta}$ that predicts the residual with respect to the ground truth. This 2.5D approach incorporates spatial context across slices while supervising the center slice reconstruction.

\subsection{{Hyperparameters}}

{Both the 2D PnP algorithm~\cite{pramanik2025fast} and our proposed 2.5D PnP algorithm only require setting three hyperparameters before reconstruction --- the total number of iterations, the number of CG steps for the data-fitting step, and the candidate values for $\beta$. We use the same values for our hyperparameters as proposed in the 2D PnP paper. Namely, we set the number of total iterations to 3, the number of CG steps to 10, and the candidates for $\beta$ to $\{2^{1-i}\}_{i=0}^{14}$.} 

\subsection{Synthetic Datasets}

We perform experiments on simulated XCT scans of Aluminum additively manufactured parts generated using Computer-Aided Design (CAD) models. 
Namely, we generate one training set that we use to train a 2D and 2.5D artifact reduction CNN, as well as two test sets.
The InD test set contains scans that match the training set in the material of the part, number of views, presence of beam hardening (BH), and noise level.
The OOD test set contains scans with different noise level or number of views than the training set.

To simulate realistic noise in cone-beam XCT projections, we apply a Gaussian approximation of the Poisson distribution. 
In this approximation, we add zero-mean Gaussian noise scaled by the square root of the signal intensity and a user-defined noise parameter $\sigma$ to the projection data; we refer to the noise parameter $\sigma$ as the ``noise level''. Let $W$ represent the ideal photon count data, obtained by forward-projecting a digital phantom (e.g., derived from a CAD model) onto a virtual detector using cone-beam geometry across multiple view angles. The noisy projection data is then expressed as:
 
\begin{equation}
W_{\text{noisy}} = W + \sqrt{W} \cdot \sigma \cdot \mathcal{N}(0, 1),
\end{equation}
where $W_{\text{noisy}}$ is the simulated noisy projection, $\sigma$ controls the noise level, and $\mathcal{N}(0, 1)$ denotes element-wise independent standard normal random variables. 
This formulation approximates Poisson statistics by leveraging the variance-to-mean relationship inherent in photon-counting statistics, and is commonly used in CT simulation pipelines when actual photon statistics are not available or when computational efficiency is prioritized.
 
In our simulations, we vary both the noise level $\sigma$ and the number of projection views to study their impact on reconstruction quality. The training phantom is forward-projected using cone-beam geometry under both full-scan (covering $360^\circ$) and short-scan (covering $197^\circ$ with a $17^\circ$ fan angle)~\cite{parker1982optimal} acquisition protocols to evaluate performance under different sampling conditions.
A short-scan is a type of scan that only measures projections at $180^\circ$ plus a small fan-angle as opposed to a full $360^\circ$ degree scan.
It was proposed originally by Parker~\cite{parker1982optimal} using Tuy's conditions~\cite{Tuy1,Tuy2} which establish that a short-scan is sufficient to get the same result as a full scan for an object fully contained in the field of view.

Table~\ref{tab:ct_data} gives an overview of the paired input and reference scans in the training and test sets, including the number of views, whether it is a short-scan, the noise level simulated, and whether the scan contains BH artifacts.

BH is a common artifact in XCT imaging caused by the poly-chromatic nature of the X-ray beam. 
As X-rays pass through dense materials, lower-energy photons are absorbed more rapidly than higher-energy photons. 
This results in a shift toward higher average photon energy (“beam hardening”) and causes nonlinear attenuation effects that lead to artifacts in the reconstructed image, such as cupping and streaks. 
To simulate data without BH, we assume the X-ray source is a single energy beam, whereas to simulate data with BH, we consider the spectrum of the X-ray source at different energies.
To mitigate BH artifacts, one typically performs a pre-processing correction step on the raw projection data. 
However, in our proposed 2.5D PnP method, the CNN model implicitly corrects for both BH artifacts and noise, which avoids the need for pre-processing. 
Thus, all the input scans in our training and test sets contain BH.

The simulated detector size is set to $1456 \times 1840$ pixels, with each pixel measuring $0.127$mm$\times 0.127$mm, matching a standard detector being used in commercial industrial XCT systems (e.g. Zeiss Metrotom). 
We use python's spekpy package (\cite{spekpy1, spekpy2}), to simulate the XCT spectrum with a peak voltage of 180kV and a 2mm Al filter typically used as pre-filter to reduce the BH effect. 
Since the detector has 1840 channels, a common rule of thumb would require 1840 views at sufficiently high signal-to-noise ratio to guarantee high-quality reconstructions when using traditional algorithms (e.g. FDK).
However, to reduce scan time and cost, as well as to increase throughput, the input scans in our dataset have significantly fewer views, in some cases by a factor of more than 10.
The reconstructed volumes have a voxel size of 17.28 $\mu$m.

\section{Results}
In this section, we summarize results demonstrating the generalizability of our proposed approach on synthetic data that is in- and out-of-distribution with respect to the training set, as well as on experimental XCT scans.

\subsection{In-Distribution (InD) Test Data}

\begin{figure*}
    \centering
    \includegraphics[width=\textwidth]{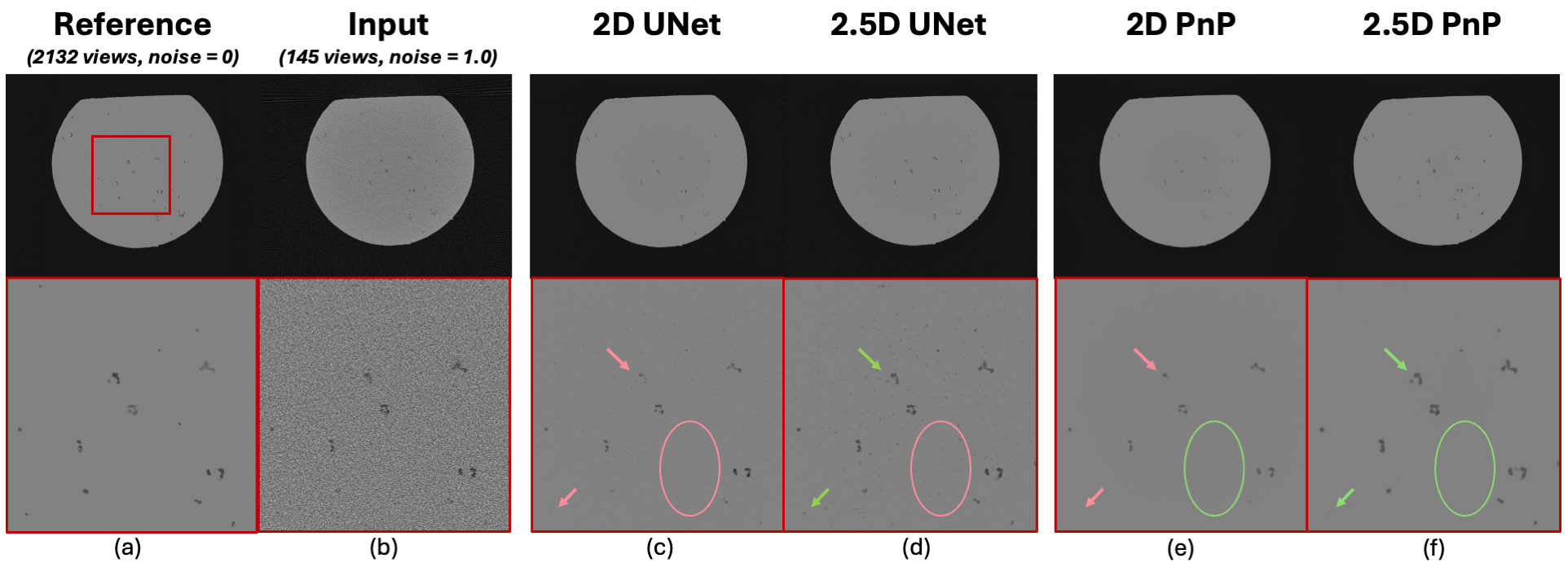}
    \caption{Comparison of (a) reference, (b) input, (c) 2D UNet, (d) 2.5D UNet, (e) 2D PnP, and (f) our proposed 2.5D PnP using a short and sparse scan with 145 views and noise level 1, which is InD with the training data. 
    2D and 2.5D PnP reduce noise in the background that 2D and 2.5D UNet do not (green and red ovals). 
    2.5D PnP generates a more accurate reconstruction of pores than 2D PnP (green and red arrows).
    }
    \label{fig:2D_vs_2.5D_PnP}
\end{figure*}

Figure~\ref{fig:2D_vs_2.5D_PnP} compares the performance of the supervised 2D UNet and 2.5D UNet models, as well as 2D PnP and our proposed 2.5D PnP using a short and sparse scan with 145 views and noise level 1.
As noted in Table~\ref{tab:ct_data}, this scan is considered InD with respect to the training data.
Both PnP methods are able to reduce the noise in the background that the UNets cannot (red and green ovals).
However, 2D PnP fails to reconstruct some small pores and distorts the shape of some large pores (red arrows), while the pores in the 2.5D PnP reconstruction closely match the reference reconstruction (green arrows).

\begin{figure}
    \centering
    \includegraphics[scale=0.65]{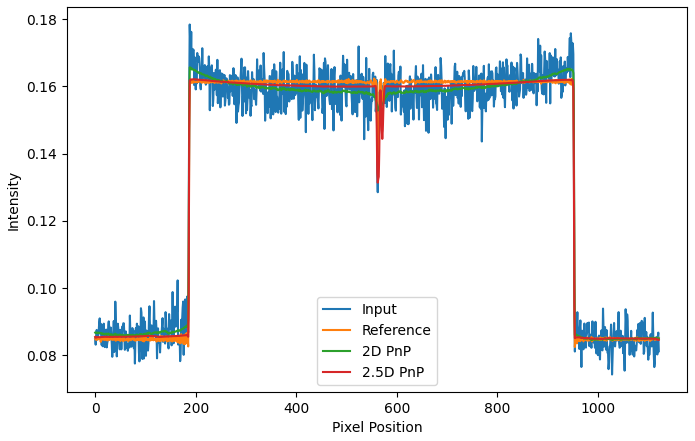}
    \caption{Comparison of pixel intensities along the center row of the center slice in the input, reference, 2D PnP, and 2.5D PnP reconstructions for a short-scan with 580 views and noise level 0.5, which is InD with the training data.
    2.5D PnP reduces cupping artifacts (i.e. having non-uniform intensity along the material with brighter edges and darker center due to BH effect) seen in the input and 2D PnP reconstructions.}
    \label{fig:line_plot}
\end{figure}

Figure~\ref{fig:line_plot} compares the pixel intensities along the center row of the center slice in the input, reference, 2D PnP, and 2.5D PnP reconstructions for a short-scan with 580 views and noise level 0.5. 
Both 2D and 2.5D PnP effectively reduce the noise from the input FDK reconstruction.
However, 2.5D PnP matches the reference pixel intensity both at the edges and the center of the reconstruction, reducing the cupping artifacts seen in both the input and 2D PnP reconstructions (i.e. having non-uniform intensity along the material with brighter edges and darker center due to the BH effect).

\begin{table}
    \centering
    \resizebox{1.0\textwidth}{!}{%
    \begin{tabular}{|c|c|c||c|c|c||c|c|c||c|c||c|c|}
        \hline \rule{0pt}{1.0\normalbaselineskip}
        Data & \# of & Noise & \multicolumn{3}{|c||}{NRMSE ($\downarrow$)} & \multicolumn{3}{|c||}{SSIM ($\uparrow$)} & \multicolumn{2}{|c||}{Recall ($\uparrow$)} & \multicolumn{2}{|c|}{Precision ($\uparrow$)}\\
            & Views & Levels & Input & 2D PnP & 2.5D PnP & Input & 2D PnP & 2.5D PnP & 2D PnP & 2.5D PnP & 2D PnP & 2.5D PnP \\
        \hline \rule{0pt}{1.0\normalbaselineskip}
         \multirow{7}{*}{InD} & 580 & 0.0 & 0.025& 0.019& \textbf{0.016} & 0.997& \textbf{0.999} & \textbf{0.999} &  0.168& \textbf{0.880}& 0.722& \textbf{0.877}\\
          & 580 & 0.5 & 0.033& 0.016& \textbf{0.010}& 0.963& 0.998&\textbf{0.999}& 0.214& \textbf{0.843}& 0.730& \textbf{0.873}\\
           & 580 & 1.0 & 0.051& 0.014& \textbf{0.009}& 0.904& 0.998& \textbf{0.999}& 0.299&\textbf{0.843}& 0.454& \textbf{0.854}\\
          & 145 & 0.0 & 0.032& 0.012& \textbf{0.011}& 0.978& \textbf{0.998}&\textbf{0.998}& 0.267& \textbf{0.782}& 0.407& \textbf{0.614}\\
          & 145 & 0.5 & 0.055& 0.012& \textbf{0.010}& 0.895& 0.998&\textbf{0.999}& 0.226& \textbf{0.794}& 0.384& \textbf{0.751}\\
          & 145 & 1.0 & 0.096& \textbf{0.016}& 0.017& 0.830&\textbf{0.998}&\textbf{0.998}& 0.248& \textbf{0.670}& 0.352& \textbf{0.719}\\
         \hline \rule{0pt}{1.0\normalbaselineskip}
          \multirow{5}{*}{OOD Noise} & 580 & 0.75 & 0.042& 0.015& \textbf{0.009}& 0.932& 0.998& \textbf{0.999}& 0.328& \textbf{0.843}& 0.467& \textbf{0.871}\\
      & 580 & 2.0 & 0.107& 0.038& \textbf{0.036}& 0.584& 0.993& \textbf{0.994}& 0.739& \textbf{0.933}& \textbf{0.757}& 0.709\\
       & 580 & 4.0 & 0.207& 0.038 & \textbf{0.036} & 0.286 & 0.992 & \textbf{0.993}& 0.737& \textbf{0.906}& \textbf{0.776}& 0.667\\
         & 145 & 0.75 & 0.075& 0.019& \textbf{0.010}& 0.856& 0.998&\textbf{0.999}& 0.290& \textbf{0.756}& 0.382& \textbf{0.747}\\
      & 145 & 2.0 & 0.211 & 0.041 & \textbf{0.040} & 0.278& 0.990 & \textbf{0.991} & 0.614& \textbf{0.831}& \textbf{0.779}& 0.686\\
      \hline\rule{0pt}{1.0\normalbaselineskip}
       \multirow{2}{*}{OOD Views} & 73 & 0.5 &  0.109& \textbf{0.041}&  0.042& 0.249& \textbf{0.978}&  0.964& 0.638& \textbf{0.879}& \textbf{0.658}& 0.306\\
      & 73 & 1.0&  0.167& \textbf{0.056}&  0.057& 0.117& \textbf{0.983}&  0.982& 0.224& \textbf{0.366}& \textbf{0.375}& 0.224\\
         \hline
    \end{tabular}
    } 
    \caption{Image quality and probability of detection metrics for InD and OOD test sets from Table~\ref{tab:ct_data}. Recall and precision are reported for flaws with diameter ranging from 75$\mu$m to 125$\mu$m. Best metrics for each testing scan are shown in bold.}
    \label{tab:quant_metrics_id}
\end{table}

\begin{figure}
    \centering
    \includegraphics[width=\linewidth]{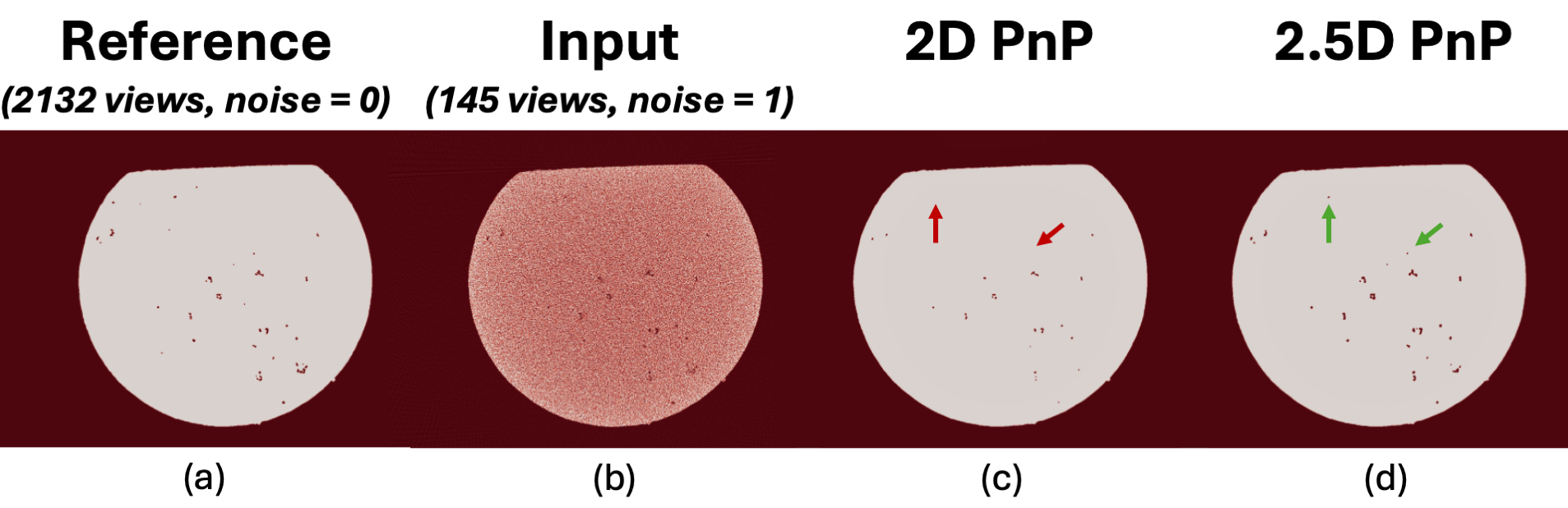}
    \caption{Comparison of Otsu thresholding segmentation of the (a) reference, (b) input, (c) 2D PnP, and (d) our proposed 2.5D PnP reconstructions for a short and sparse scan with 145 views and noise level 1, which is InD with the training data. The detected defects are shown in red overlaid on the grayscale reconstruction slice. More defects are detected in the 2.5D PnP reconstruction than the 2D PnP reconstruction (red and green arrows).}
    \label{fig:segmentation}
\end{figure}

Table~\ref{tab:quant_metrics_id} summarizes the NRMSE and structural similarity index metric (SSIM) for 2D and 2.5D PnP, showing that 2.5D PnP performs better on InD data compared to 2D PnP.
However, typical metrics such as NRMSE (or equivalently PSNR) and SSIM are not sufficient for analyzing performance when dealing with data from experimental scientific imaging applications such as industrial XCT~\cite{cai2017image}. 
In order to have useful and relevant metrics, we compare the impact of 2D PnP and our proposed 2.5D PnP on the ability to detect defects within the part using Otsu thresholding~\cite{otsu1975threshold} for segmentation in Figure~\ref{fig:segmentation}.
The detected defects are shown in red overlaid on the grayscale reconstruction slice.
More defects are detected in the 2.5D PnP reconstruction than the 2D PnP reconstruction (red and green arrows). 

\begin{figure}
        \includegraphics[width=0.8\textwidth]{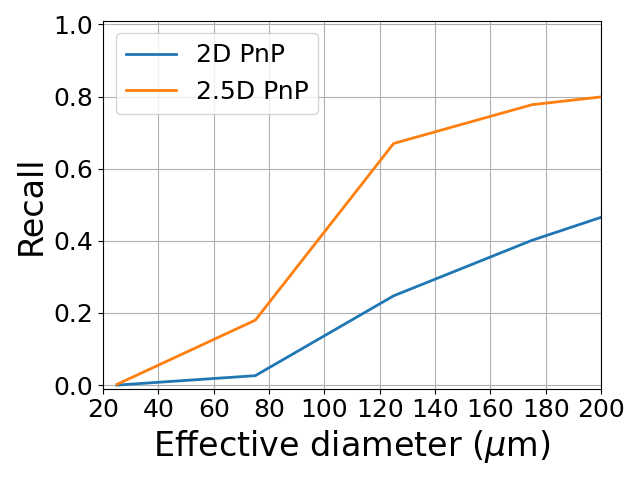} \\
        \includegraphics[width=0.8\textwidth]{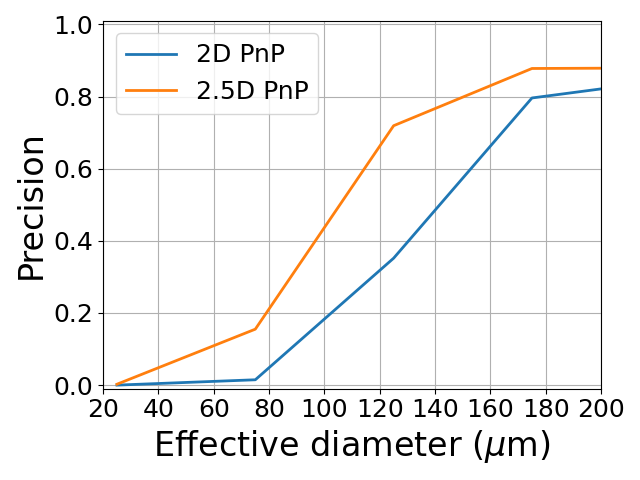}
    \caption{Comparison of recall and precision for 2D and 2.5D PnP reconstructions (blue and orange) for a short and sparse scan with 145 views and noise level 1. 2.5D PnP achieves higher recall and precision over all defect diameters.}
    \label{fig:recall_precision}
\end{figure}

Figure~\ref{fig:recall_precision} shows the recall and precision curves calculated from the 3D segmented reconstructions, using the segmented reference as ground truth. 
2.5D PnP achieves higher recall and precision over all defect diameters, supporting our conclusion that 2.5D PnP enables better detection of defects. 

In Table~\ref{tab:quant_metrics_id}, we report the NRMSE and SSIM, as well as recall and precision for flaws with diameter ranging from 75$\mu$m to 125$\mu$m in the 2D and 2.5D PnP reconstructions.
2.5D PnP achieves better image quality metrics (NRMSE and SSIM) for almost all InD testing scans, which is consistent with the qualitative results we have observed. 
Note that the SSIM for both 2D and 2.5D PnP is very close to 1, implying that both methods are able to attain high-quality features. However, 2.5D PnP achieves a lower NRMSE, implying that it is able to match the data better than 2D PnP.
We also note that 2D and 2.5D PnP significantly improve both NRMSE and SSIM as compared to the input FDK and offer more consistent performance across varying number of views and noise levels.
Additionally, 2.5D PnP consistently achieves significantly higher recall and precision for InD data, implying that it enables more accurate defect detection.

\subsection{Out-Of-Distribution (OOD) Test Data}
One of the main advantages of PnP-based methods is their ability to better generalize to OOD data. 
On the other hand, end-to-end deep learning-based methods (like UNet) need to be retrained for OOD data.
In this section, we compare the performance of 2D and 2.5D UNet and PnP on simulated scans with noise that is OOD with respect to the training set.

\begin{figure*}
     \centering
    \includegraphics[width=\textwidth]{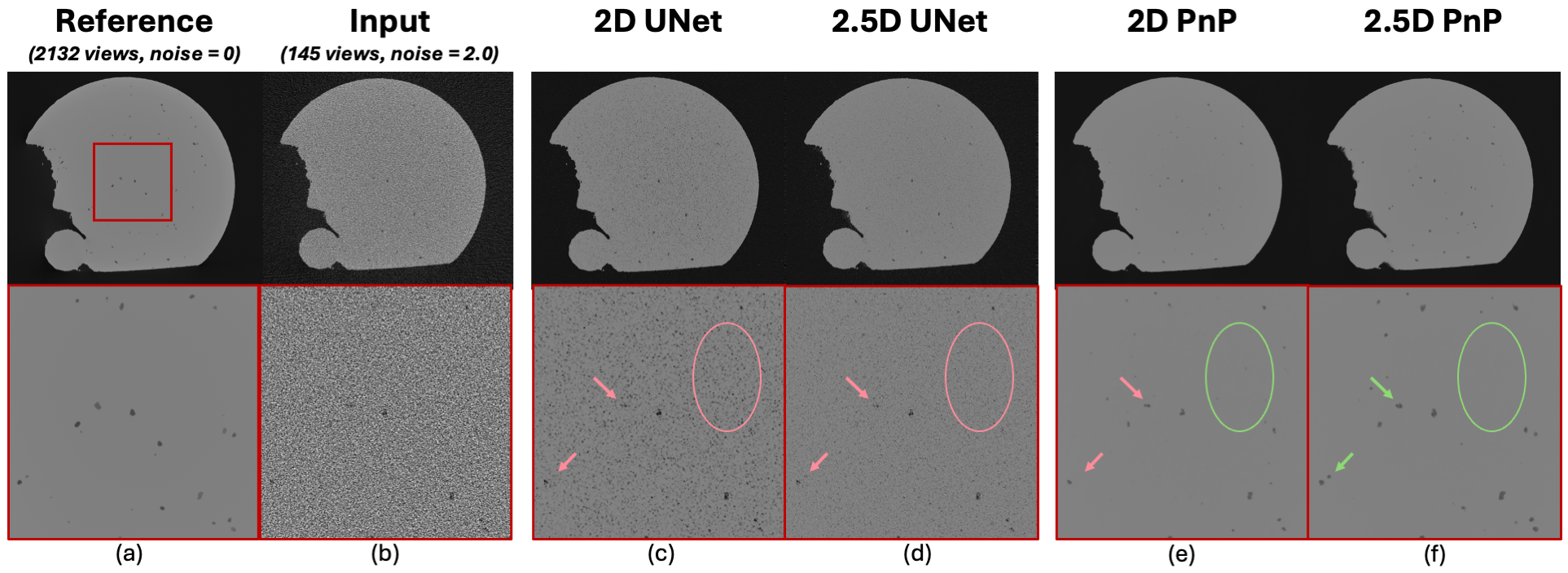}
    \caption{Comparison of (a) reference, (b) input, (c) 2D UNet, (d) 2.5D UNet, (e) 2D PnP, and (f) our proposed 2.5D PnP using a short and sparse scan with 145 views and noise level 2, which is OOD with the noise in the training data. 2D and 2.5D PnP effectively remove background noise, with 2.5D PnP best preserving pore structure.}
    \label{fig:UNet_vs_PnP_OOD_noise}
\end{figure*}

Figure~\ref{fig:UNet_vs_PnP_OOD_noise} shows a comparison of 2D UNet, 2.5D UNet, 2D PnP, and our proposed 2.5D PnP using a short and sparse scan with 145 views and noise level 2, which is noisier than the training data.
Both 2D and 2.5D PnP are able to reduce the noise in the background (green ovals), while 2D and 2.5D UNet are unable to fully remove the noise (red ovals).
Additionally, both 2D UNet and 2D PnP are unable to reconstruct some small pores and alter the shape of some of the larger pores (red arrows), while 2.5D UNet and 2.5D PnP are able to more accurately reconstruct these pores (green arrows).

\begin{figure}
    \centering
    \includegraphics[width=\linewidth]{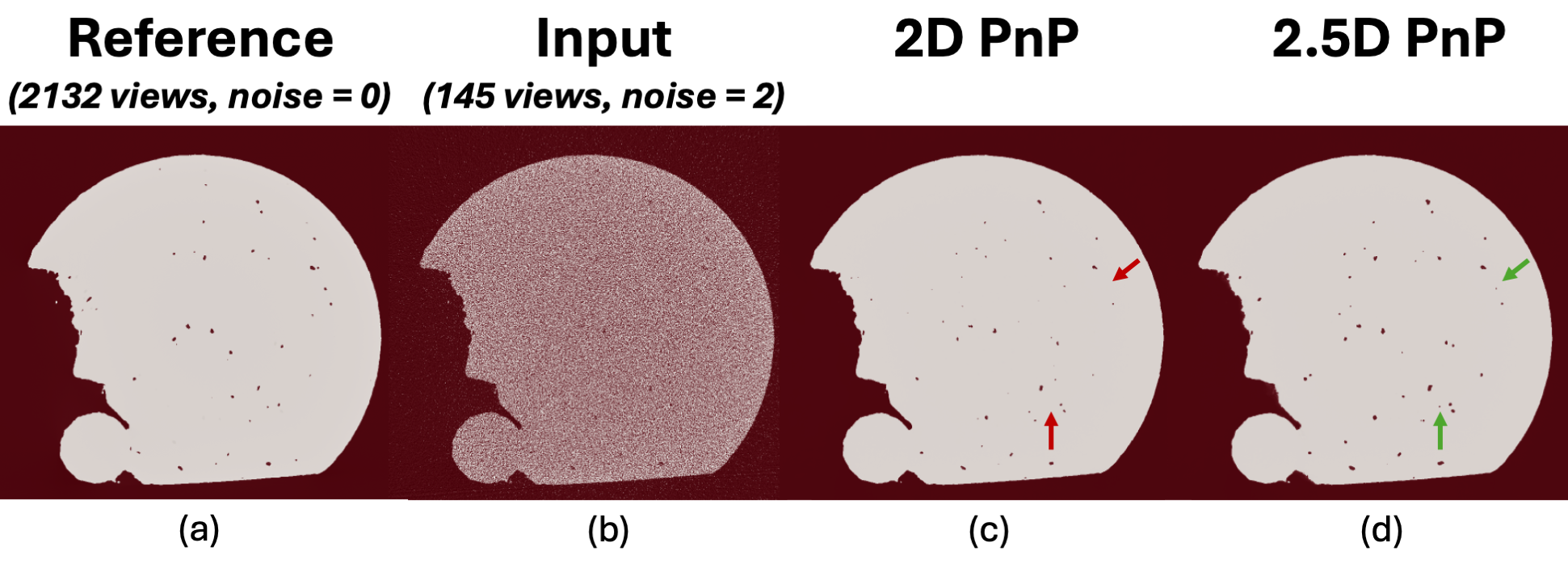}
    \caption{Comparison of Otsu thresholding segmentation of the (a) reference, (b) input, (c) 2D PnP, and (d) 2.5D PnP reconstructions for a short and sparse scan with 145 views and noise level 2, which is OOD with the noise in the training data. 2D PnP fails to detect defects that 2.5D PnP is able to detect (red and green arrows).}
    \label{fig:detection_OOD_noise}
\end{figure}

\begin{figure}
            \includegraphics[width=0.8\textwidth]{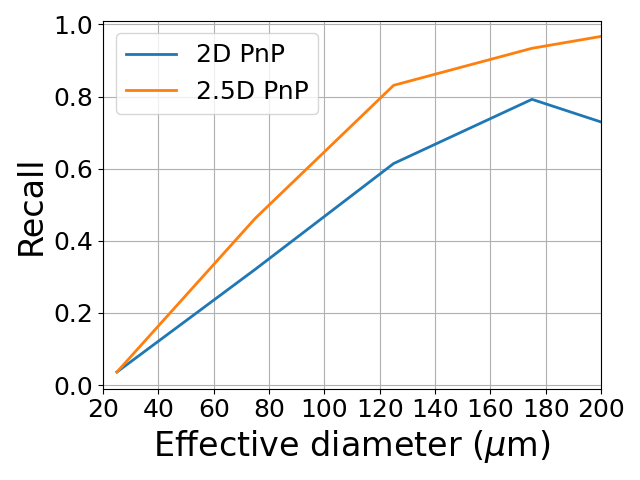} \\
            \includegraphics[width=0.8\textwidth]{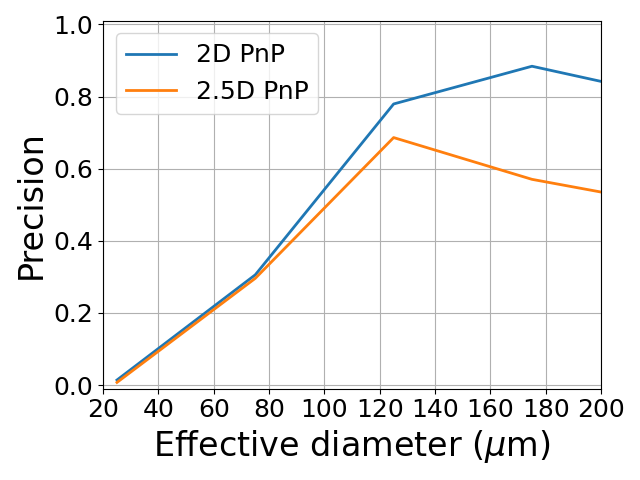}
    \caption{Comparison of recall and precision for 2D and 2.5D PnP reconstructions, blue and orange respectively, using 145 views and noise level 2. 2.5D PnP achieves higher recall but lower precision than 2D PnP, indicating better pore detection at the cost of some false positives.}
    \label{fig:recall_precision_OOD_noise}
\end{figure}

Figure~\ref{fig:detection_OOD_noise} compares the impact of 2D and 2.5D PnP on the detected defects from a short and sparse scan with 145 views and noise level 2, using Otsu thresholding for segmentation. 
The detected defects are shown in red overlaid on the grayscale reconstruction slice.
2D PnP fails to detect defects that 2.5D PnP is able to detect (red and green arrows).
We observe a similar pattern in the recall and precision curves shown in Figure~\ref{fig:recall_precision_OOD_noise}. 
2.5D PnP achieves a significantly higher recall, implying that it can better detect pores. 
2D and 2.5D PnP have similar precision for smaller pores.
However, 2D PnP has better precision for larger pores, implying that 2.5D PnP detects more false positives for larger pores.

In Table~\ref{tab:quant_metrics_id}, we report the NRMSE and SSIM, as well as recall and precision for flaws with diameter ranging from 75$\mu$m to 125$\mu$m in the 2D and 2.5D PnP reconstructions for scans with OOD noise levels.
For scans with noise level in between training noise levels (i.e. noise level 0.75), 2.5D PnP achieves better image quality metrics and better recall and precision.
For scans with higher noise levels (i.e. noise levels 2.0 and 4.0), 2.5D PnP achieves better image quality metrics and better recall, but lower precision. This implies that 2.5D PnP enables detection of more defects at the cost of slightly more false positives when reconstructing scans with more noise than the training data.

\section{Experimental Data}
In addition to synthetic data with OOD noise, our proposed PnP method performs well on experimental XCT scans, even though the artifact reduction prior is trained only on the synthetic training set from Table~\ref{tab:ct_data}. 
In this section, we compare the performance of 2D and 2.5D PnP on parts made of aluminum-cerium (Al-Ce) with a short and sparse scan consisting of 145 views with 180 kV source voltage and 8s integration time. 
For reference, we use an MBIR reconstruction of a short-scan with 580 views. 
{The reconstruction is of size $1356\times 1356\times 1264$. Importantly, the scalability of our method enables processing volumes of this large volume size, which many existing approaches struggle to handle efficiently or practically~\cite{hofmann2022principles}.}

\begin{figure}
    \centering
    \includegraphics[width=\linewidth]{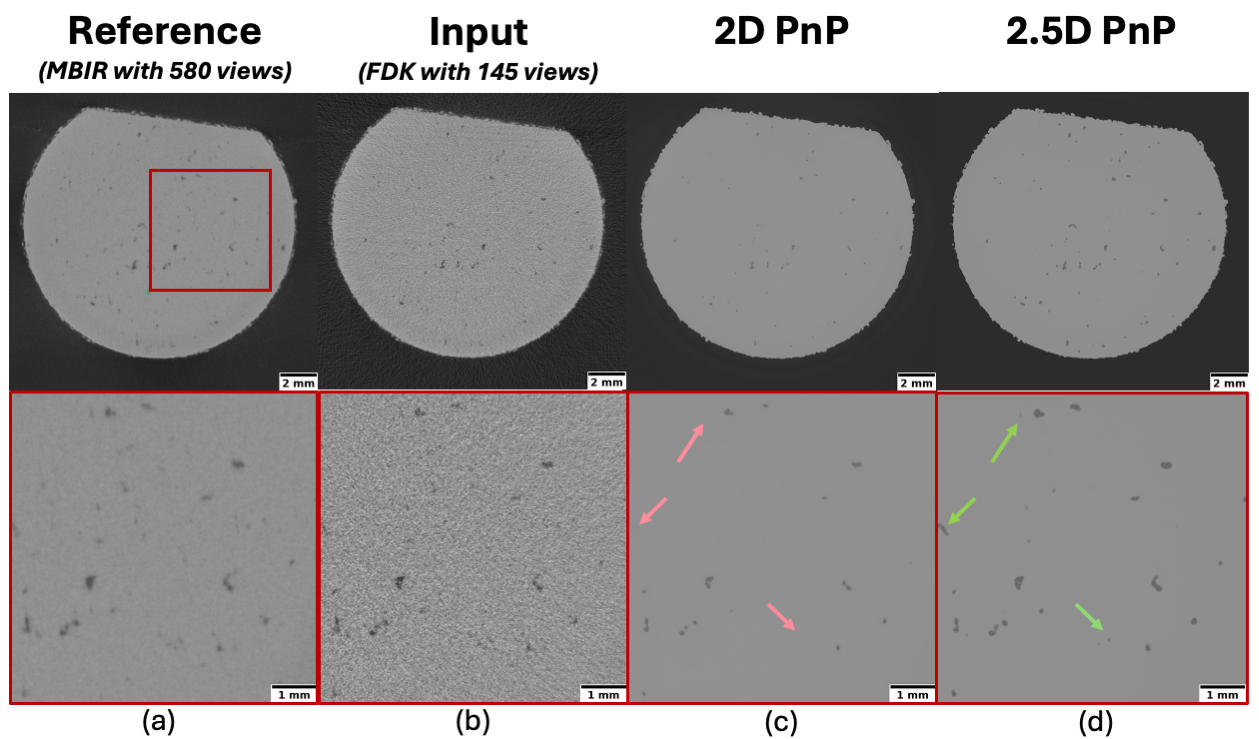}
    \caption{Comparison of (a) reference, (b) input, (c) 2D PnP, and (d) proposed 2.5D PnP on experimental XCT short and sparse scan of Al-Ce part with 145 views. Both PnP reconstructions reduce noise compared to FDK; however, 2.5D PnP recovers more pores (red and green arrows).}
    \label{fig:PnP_real}
\end{figure}

Figure~\ref{fig:PnP_real} compares the reference MBIR, input FDK, 2D PnP, and proposed 2.5D PnP reconstructions. 
Both PnP reconstructions contain less noise with more distinguishable pores than the FDK reconstruction.
Our proposed 2.5D PnP reconstruction preserves more pores in the part than 2D PnP (green and red arrows).

{Table~\ref{tab:metrics_real} reports the NRMSE, SSIM, signal-to-noise ratio (SNR), and contrast-to-noise ratio (CNR) of the input FDK, 2D PnP, and 2.5D PnP reconstructions of a 145 view scan with respect to the MBIR reconstruction from a 580 view scan.
To compute the SNR and CNR, we select two $50\times 50$ regions within 10 slices of the reconstructions that contain either only background or only material (no defects). Then, we compute the SNR (in dB) and CNR (unitless) as
\begin{equation}
    SNR = 20 \log_{10} \left(\frac{\mu_{\text{material}}}{\sigma_{\text{material}}} \right) 
\end{equation}
\begin{equation}
    CNR = \frac{\left | \mu_{\text{background}} - \mu_{\text{material} } \right |}{\sqrt{\sigma_{\text{background}}^2 +\sigma_{\text{material} } ^2}} 
\end{equation}
where $\mu_{\text{background}}$ and $\sigma_{\text{background}}$ are the mean and standard deviation over the background region and $\mu_{\text{material}}$ and $\sigma_{\text{material}}$ are the mean and standard deviation over the material region. 
2D and 2.5D PnP attain higher SNR and CNR than both FDK and MBIR, with 2.5D PnP achieving the highest values. 
Considering that the proposed approaches significantly suppress the noise both in the material and background regions, the very high CNR and SNR values are expected. 
Additionally, both 2D and 2.5D PnP significantly outperform the input FDK, attaining similar NRMSE and SSIM values. 
2D PnP attains slightly better image quality metrics; however, as shown in Table~\ref{tab:quant_metrics_id} this does not translate to the task-specific metric of defect detection. }

\begin{table}
    \centering
    \begin{tabular}{|c|c|c|c|c|}
    \hline \rule{0pt}{1.0\normalbaselineskip}
        Image Quality Metric & Reference MBIR & Input FDK & 2D PnP & 2.5D PnP \\
        \hline \rule{0pt}{1.0\normalbaselineskip}
         NRMSE (unitless) & -- & 0.179 & \textbf{0.148} & 0.152 \\
         SSIM (unitless) & -- & 0.916 & \textbf{0.991} & 0.990 \\
         SNR (in dB; ($\mu$, $\sigma$)) & (33.03, 0.61) & (17.51, 0.31) & (51.13, 1.52) & (\textbf{62.49}, 5.68 )\\
         CNR (unitless; ($\mu$, $\sigma$)) & (13.34, 0.14) & (4.75, 0.11) & (298.15, 36.16) & (\textbf{1326.21}, 493.47) \\
         \hline
    \end{tabular}
    \caption{Image quality metrics for experimental XCT short and sparse scan of Al-Ce part with 145 views, using MBIR with 580 views as reference. The best results are shown in bold. The PnP methods significantly outperform the input FDK, with 2.5D PnP attaining the best SNR \& CNR. 2D PnP attains slightly better NRMSE \& SSIM; however, this does not translate to the task-specific metric of defect detection.}
    \label{tab:metrics_real}
\end{table}

\begin{figure}[!htbp]
    \centering
    \includegraphics[width=\linewidth]{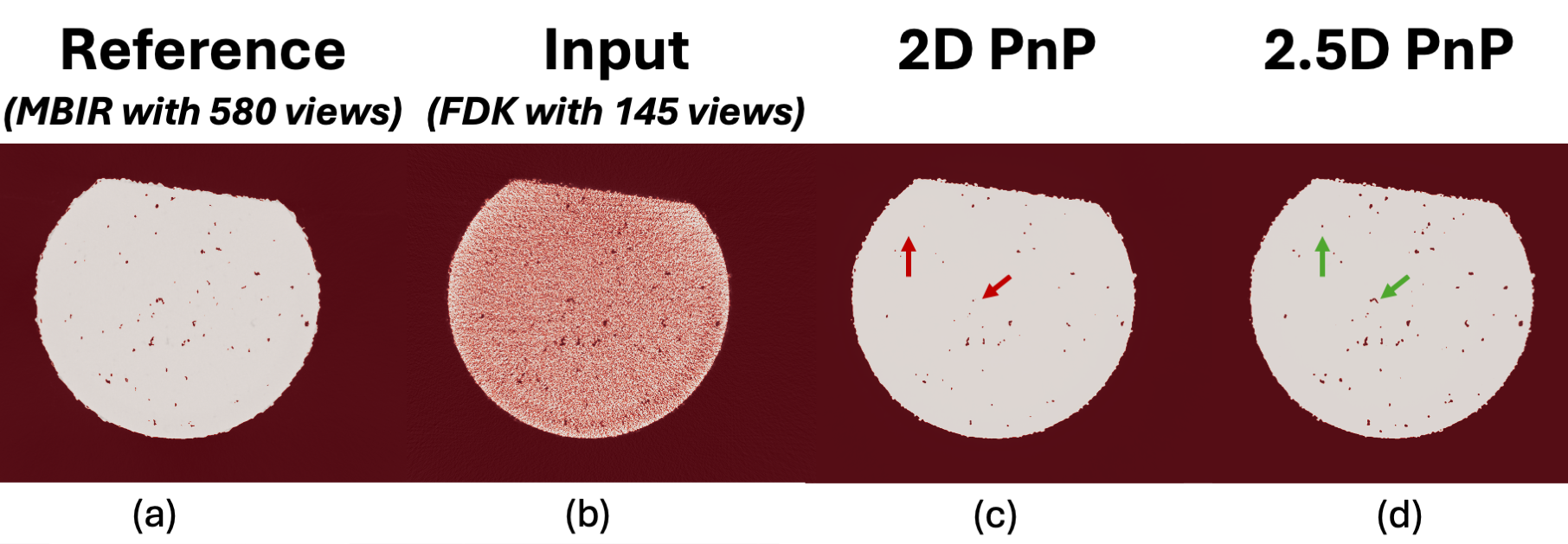}
    \caption{Comparison of Otsu thresholding segmentation of the (a) reference, (b) input, (c) 2D PnP, and (d) 2.5D PnP reconstructions for experimental short and sparse scan of Al-Ce part with 145 views. The detected defects are shown in red overlaid on the grayscale reconstruction slice. 2.5D PnP enables the detection of more defects than 2D PnP. 
    }
    \label{fig:defects_real}
\end{figure}

\begin{figure}
    \includegraphics[width=0.7\textwidth]{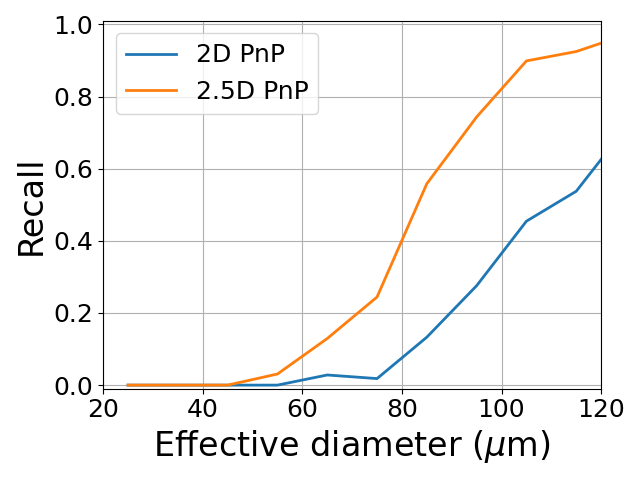} \\
    \includegraphics[width=0.7\textwidth]{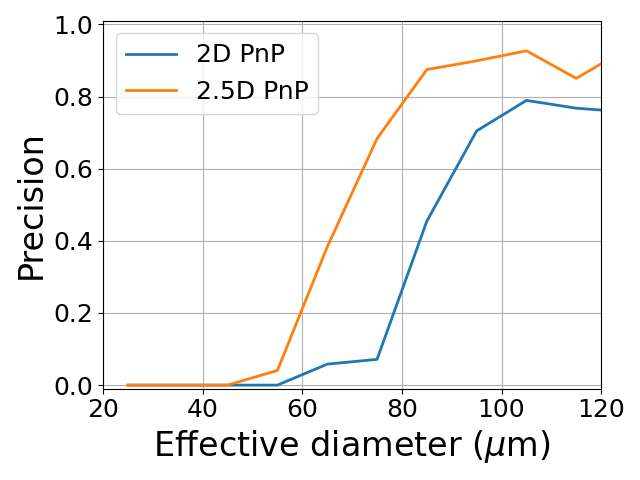}
    \caption{Comparison of recall and precision for 2D and 2.5D PnP reconstructions, blue and orange respectively, using experimental short and sparse scan of Al-Ce part with 145 views. 2.5D PnP achieves higher recall and precision over all defect diameters. }
    \label{fig:recall_precision_real}
\end{figure}

Figure~\ref{fig:defects_real} compares the impact of 2D and 2.5D PnP on the ability to detect defects within the part, using Otsu thresholding for segmentation. 
The detected defects are shown in red overlaid on the grayscale reconstruction slice.
2.5D PnP detects more pores within the part and better preserves the shape and size of the detected pores (red and green arrows). 
This observation is further justified by the recall and precision curves shown in Figure~\ref{fig:recall_precision_real}. 
2.5D PnP achieves significantly higher recall and precision across all defect diameters, supporting our conclusion that 2.5D PnP enables significantly better defect detection, even when reconstructing experimental XCT scans using a prior trained only on synthetic data.
{It should be emphasized that we chose Otsu thresholding~\cite{otsu1975threshold} since it is a simple, parameter-free, and widely-used approach. It avoids tuning across reconstruction qualities and allows for consistent, reproducible evaluation across methods. While CNN-based segmentation approaches could improve segmentation performance, they also introduce confounding factors like retraining and hyperparameter sensitivity. Our focus was to assess whether improved image quality translates into improved detectability using a fixed, minimal segmentation baseline, which can easily be achieved using Otsu thresholding.}

{Table~\ref{tab:timing_real} compares the reconstruction time and peak memory usage for 2D PnP and proposed 2.5D PnP when reconstructing an XCT volume of size $1356 \times 1356 \times 1264$ using four Nvidia H100 GPUs with 80 GB of memory. Both the 2D and 2.5D PnP reconstructions take approximately 48 minutes to complete and require approximately 35 GB of GPU memory during processing. While this is more expensive than FDK---which takes approximately 1 minute and uses approximately 0.5 GB of GPU memory---both PnP methods produce significantly higher-quality reconstructions that substantially improve artifact suppression and defect detection. Importantly, 2.5D PnP is also an order of magnitude less expensive than MBIR, which requires approximately 6 hours per volume and uses approximately 300 GB of GPU memory, while delivering comparably high-quality results. Moreover, our 2.5D PnP method is only slightly more expensive than 2D PnP ($\approx 2$ seconds and $\approx 1$ GB), but provides noticeably improved reconstruction fidelity and higher probability of detection for key features. Thus, 2.5D PnP offers a compelling balance between computational cost and reconstruction quality.}

\begin{table}
    \centering
    \begin{tabular}{|c|c|c|}
    \hline \rule{0pt}{1.0\normalbaselineskip}
        Method & Time (secs) & Peak Memory Usage (MB) \\
        \hline \rule{0pt}{1.0\normalbaselineskip}
         2D PnP & 2,924.24 & 34,922.06 \\
         2.5D PnP & 2,926.28 & 35,895.36 \\
         \hline
    \end{tabular}
    \caption{Runtime (in seconds) and peak GPU memory usage (in MB) for each reconstruction method. Our proposed 2.5D PnP is much less expensive than MBIR, which requires approximately 6 hours per volume and uses approximately 300 GB of GPU memory, while delivering comparably high-quality results.}
    \label{tab:timing_real}
\end{table}

\section{Discussion}
Our synthetic OOD test set also includes two scans with fewer views than the training set. Namely, we test on short and sparse scans with 73 views, which are very sparse and pose a difficult challenge for reconstruction algorithms.
Table~\ref{tab:quant_metrics_id} reports the image quality and defect detection metrics for 2D and 2.5D PnP applied to these short and sparse scans with a noise level of 0.5 and 1.0.
{Despite being trained on denser view distributions, both 2D and 2.5D PnP frameworks demonstrate strong generalization capabilities to this OOD data. The 2D PnP consistently achieves higher image quality scores and improved precision, whereas the 2.5D PnP shows increased recall, detecting a larger number of defects but with somewhat reduced precision. This trade-off suggests that 2.5D PnP is more sensitive in identifying subtle defect features, likely due to its exploitation of inter-slice contextual information.}

{We hypothesize that the increase in false positives for 2.5D PnP arises from “pore-like” structures induced by view sparsity artifacts that appear consistently across neighboring slices. Nonetheless, the ability of 2.5D PnP to leverage volumetric information highlights its scalability and adaptability when applied to diverse scan protocols and noise conditions outside the original training distribution.}

\section{Conclusion}

In this paper, we presented a PnP framework with a 2.5D artifact reduction prior for industrial computed tomography of large scale 3D data.
The proposed approach improves XCT reconstruction quality and defect detection in case studies from additively manufactured parts. 
We also demonstrated that 2.5D PnP can effectively suppress BH artifacts during reconstruction, eliminating the need for separate pre-processing. 
Specifically, our proposed method produces reconstructions with internal pores that more closely match the ground truth, even under OOD noise. 
In contrast, using a 2D prior leads to distortions in pore shape, size, and contrast --- hindering accurate defect detection.
Notably, we observed strong performance on experimental XCT data using a 2.5D prior trained entirely on synthetic data, highlighting our method’s ability to generalize across domains.

{Importantly, our results underscore the robustness and scalability of the proposed 2.5D PnP framework for real-world applications where scan parameters may vary substantially. The demonstrated generalization to OOD scans with differing noise levels suggests that our approach can be effectively scaled and adapted OOD imaging scenarios without retraining or extensive parameter tuning.}

While our method performs well under a range of conditions, it currently struggles to generalize to OOD view sparsity. In future work, we aim to address this limitation by improving the robustness of the learned prior and refining the tuning of reconstruction parameters. 
We also plan to perform more thorough testing on experimental XCT scans and extend our testing to a wider range of OOD scenarios, including denser materials. 

\bibliography{refs}

\end{document}